\PassOptionsToPackage{dvipsnames}{xcolor}

\documentclass[10pt,preprint,twocolumn]{aastex631}

\def\macc   {$\dot{M}_{\rm acc}$}
\def\lacc   {$L_{\rm acc}$}

\def\halpha  {H$\alpha$}

\def\msun {$M_{\odot}$}
\def\rsun {$R_{\odot}$}
\def\lsun {$L_{\odot}$}

\newcommand{\vsini}{$v\sin i$}

\newcommand{\uat}[2]{\href{http://vocabs.ands.org.au/repository/api/lda/aas/the-unified-astronomy-thesaurus/current/resource.html?uri=http://astrothesaurus.org/uat/#1}{#2 (#1)}}

\shorttitle{PDS\,70 inner disc wind}
\shortauthors{J. Campbell-White et al.}

\graphicspath{{./}{figures/}}

\begin{document}

\title{A magnetically driven disc wind in the inner disc of PDS 70\footnote{Based on observations collected at the European Southern Observatory under ESO programmes  098.C-0739, 0104.C-0418, 105.205R, 106.20Z8 and 1101.C-0557}}

\email{jcampbel@eso.org}

\author[0000-0002-3913-3746]{Justyn Campbell-White}
\affiliation{European Southern Observatory, Karl-Schwarzschild-Strasse 2, 85748 Garching bei M\"unchen, Germany}

\author[0000-0003-3562-262X]{Carlo F. Manara}
\affiliation{European Southern Observatory, Karl-Schwarzschild-Strasse 2, 85748 Garching bei M\"unchen, Germany}

\author[0000-0002-7695-7605]{Myriam Benisty}
\affiliation{Universit\'{e} C\^{o}te d'Azur, Observatoire de la C\^{o}te d'Azur, CNRS, Laboratoire Lagrange, F-06304 Nice, France}
\affiliation{Universit\'{e} Grenoble Alpes, CNRS, Institut de Plan\'{e}tologie et d'Astrophysique (IPAG), F-38000 Grenoble, France}

\author{Antonella Natta}
\affiliation{Dublin Institute for Advanced Studies, 31 Fitzwilliams Place, Dublin, Ireland}

\author[0000-0001-8194-4238]{Rik A. B. Claes}
\affiliation{European Southern Observatory, Karl-Schwarzschild-Strasse 2, 85748 Garching bei M\"unchen, Germany}

\author[0000-0002-0474-0896]{Antonio Frasca}
\affiliation{INAF - Osservatorio Astrofisico di Catania, via S. Sofia, 78, 95123 Catania, Italy}

\author[0000-0001-7258-770X]{Jaehan Bae}
\affiliation{Department of Astronomy, University of Florida, Gainesville, FL 32611, USA}

\author[0000-0003-4689-2684]{Stefano Facchini}
\affiliation{Dipartimento di Fisica, Universit\'{a} degli Studi di Milano, via Celoria 16, 20133 Milano, Italy}

\author[0000-0002-0786-7307]{Andrea Isella}
\affiliation{Department of Physics and Astronomy, Rice University, 
6100 Main Street, MS-108,
Houston, TX 77005, USA}

\author[0000-0002-1199-9564]{Laura P\'erez}
\affiliation{Departamento de Astronom\'ia, Universidad de Chile, Camino El Observatorio 1515, Las Condes, Santiago, Chile}

\author[0000-0001-8764-1780]{Paola Pinilla}
\affiliation{Mullard Space Science Laboratory, University College London, Holmbury St Mary, Dorking, Surrey RH5 6NT, UK}

\author[0000-0002-8421-0851]{Aurora Sicilia-Aguilar}
\affiliation{SUPA, School of Science and Engineering, University of Dundee, Nethergate, DD1 4HN, Dundee, UK}

\author[0000-0003-1534-5186]{Richard Teague}
\affiliation{Department of Earth, Atmospheric, and Planetary Sciences, Massachusetts Institute of Technology, Cambridge, MA 02139, USA}

\begin{abstract}

PDS\,70 is so far the only young disc where multiple planets have been detected by direct imaging. The disc has a large cavity when seen at sub-mm and NIR wavelengths, which hosts two massive planets. This makes PDS\,70 the ideal target to study the physical conditions in a strongly depleted inner disc shaped by two giant planets, and in particular to test whether disc winds can play a significant role in its evolution. 
Using X-Shooter and HARPS spectra, we detected for the first time the wind-tracing [O\,I] 6300\AA\ line, and confirm the low-moderate value of mass-accretion rate in the literature.
The [O\,I] line luminosity is high with respect to the accretion luminosity when compared to a large sample of discs with cavities in nearby star-forming regions. The FWHM and blue-shifted peak of the [O\,I] line suggest an emission in a region very close to the star, favouring a magnetically driven wind as the origin.
We also detect wind emission and high variability in the He I 10830\AA\ line, which is unusual for low-accretors. We discuss that, although the cavity of PDS\,70 was clearly carved out by the giant planets, the substantial inner disc wind could also have had a significant contribution to clearing the inner-disc. 

\end{abstract}


%

\section{Introduction}
\label{sec:intro}

The search for planets around young stellar objects (YSOs) during the planet formation stage is still an ongoing challenge in astronomy. So is the cause and evolution of the various disc substructures that are now ubiquitously observed around YSOs \citep[e.g.,][]{2020ARA&A..58..483A,2022arXiv220309991B}. Although it could be assumed that such substructures are exclusively the direct result of embedded protoplanets, this assumption does not match with the observed exo-planet population \citep[e.g.,][]{2019MNRAS.486..453L}.  

Whilst the simple reasoning for disc substructures is pressure perturbations in the disc, the physical source can theoretically be explained not only by planets, but also by thermal, magneto-hydrodynamical and gravitational fluid instabilities within the disc material \citep[for review, see][]{2022arXiv221013314B}. Cavities in the inner disc caused by substantial mass loss may be the result of magneto-hydrodynamical (MHD)  \citep{2018ApJ...865..102T} or photoevaporative winds \citep{2014prpl.conf..475A}. Consecutive gaps in the discs could also be the result of magnetic field concentrations within the disc, inherent in MHD zonal flows \citep{2017ApJ...835..230F,2019A&A...625A.108R,2020A&A...639A..95R}. Photoevaporation could be responsible for inner disc depletion \citep[e.g.,][]{2021A&A...655A..18G}, however, this effect alone may not be strong enough to form the large cavities we observe in some transition discs \citep[TDs,][]{2016PASA...33....5O,2019MNRAS.487..691P}. 

In order to observationally disentangle the influence of disc winds on the creation of disc substructures, we should first understand whether there is any traces of a disc wind in systems where we are certain the large cavity is due to the presence of protoplanets. We may then be able to look to how such disc-winds can differ or enhance the effects of substructure formation due to embedded planets \citep[e.g.,][]{2023ApJ...946....5A,2023arXiv230511784W}.

PDS\,70, a young \citep[5.4 Myr,][]{2016MNRAS.461..794P}, nearby \citep[$\sim$112 pc,][]{2021A&A...649A...1G} star is so far the only system hosting multiple directly imaged forming planets. A large cavity hosting two massive protoplanets has been observed from sub-mm to NIR observations \cite[e.g.,][]{2018A&A...617A..44K,2019NatAs...3..749H,2021ApJ...916L...2B}. For the central star, \halpha\ equivalent width (EW) and UV flux measurements classed PDS\,70 as a non-accreting weak-line T-Tauri star \citep[WTTS,][]{2002MNRAS.336..197G,2020MNRAS.491L..56J}. However, further analysis with magnetospheric modelling of the \halpha\ line, accounting for chromospheric contributions, revealed a variable low- to moderate-accretion rate of 0.6 -- 2.2  $\times 10 ^{-10}$\msun/year \citep{2020ApJ...892...81T} with an inverse P-Cygni profile appearing and disappearing with the same periodicity as the stellar rotation. Observations with the Space Telescope Imaging Spectrograph (STIS) of the Hubble Space Telescope (\textit{HST}) confirmed both a significant chromospheric contribution from various UV emission lines and similar accretion rate with measurement of the accretion-sensitive C IV line \citep{2022ApJ...938..134S}. They also revealed the presence of fluorescent H$_2$ in the UV spectra, which is unusual for WTTS and would be pumped by Ly$\alpha$ emission. \citet{2020ApJ...892...81T} also presented a low-resolution He I 10830\AA\ profile that is indicative of both accretion and a wind. 

The observed X-ray and ultraviolet (XUV) luminosity of PDS\,70 suggests there should be a photoevaporative mass loss driven by this ionising radiation, which is potentially observable by disc-wind tracers. This was suggested by \citet{2020MNRAS.491L..56J} from analysis of the \textit{SWIFT} observations, where they predict a mass loss rate  $\sim 10 ^{-8}$\msun/yr. These XUV measurements were later confirmed with follow-up \textit{XMM-Newton} observations in \citet{2023MNRAS.519.4514J}. Given this substantial mass loss rate, the disc would be dispersed in less than $\sim$ 1 Myr. However, this is for a total disc mass of $10^{-2} - 10^{-3}$\msun\ \citep{2018A&A...617A..44K}, assuming the typical dust-to-gas ratio of 100, and that the photoionisation is efficient in reaching the outer disc. It may be that the inner-disc is more optically thick than previously thought, as shown by \citet{2021ApJ...916L...2B}, which would not allow all of the XUV flux to reach the outer-disc. Study of the \textit{HST} STIS spectra also suggested such photoevaporation only impacts the inner disc, with the planets and outer disc shielded until the inner disc is dissipated \citep{2022ApJ...938..134S}. This study also predicts a more modest mass loss rate on the order of  $\sim 10 ^{-10}$\msun/year, based on the C IV luminosity. 

If such photoevaporation or an MHD disc wind is present in the system, it may be detectable from forbidden emission lines. It is well established that such emission is a direct tracer of out-flowing material from the star and disc \citep{2022arXiv220310068P}, be it high-velocity jets \citep[e.g.,][]{1995ApJ...452..736H,2018A&A...609A..87N} or lower velocity disc winds \citep[e.g.,][]{2013ApJ...772...60R,2014A&A...569A...5N,2018ApJ...868...28F}. Lines such as [O\,I] 6300\,\AA\ have been spectrally resolved into velocity components, with models suggesting different physical origins \citep[e.g.,][]{2020MNRAS.496..223W}, however, it remains difficult to conclusively disentangle such origins i.e., thermal, non-thermal or magnetic \citep{2020ApJ...904L..27N}. Recent work has suggested that most, if not all of the low velocity emission should be due to MHD winds \citep[e.g.,][]{2016ApJ...831..169S,2018A&A...620A..87M,2019ApJ...870...76B,2023ApJ...945..112F}. \citet{2019ApJ...870...76B} further showed that the presence of a cavity in the disc results in no high velocity component, with narrower [O\,I] lines produced in more depleted cavities. 

In this work, we present analysis of medium- and high-resolution spectra of PDS\,70, which are detailed in Sec.\,\ref{sec:obs}. We present confirmation of the accretion rate measurements from these data in Sec.\,\ref{sec:analysis}, along with extraction of disc-wind tracing emission lines. These results are then compared to those of other Class II stars, WTTS and TDs and further discussed in Sec.\,\ref{sec:discussion}. We then report out conclusions in Sec.\,\ref{sec:conclusions}.

\begin{figure*}
\centering
	\includegraphics[width=\textwidth]{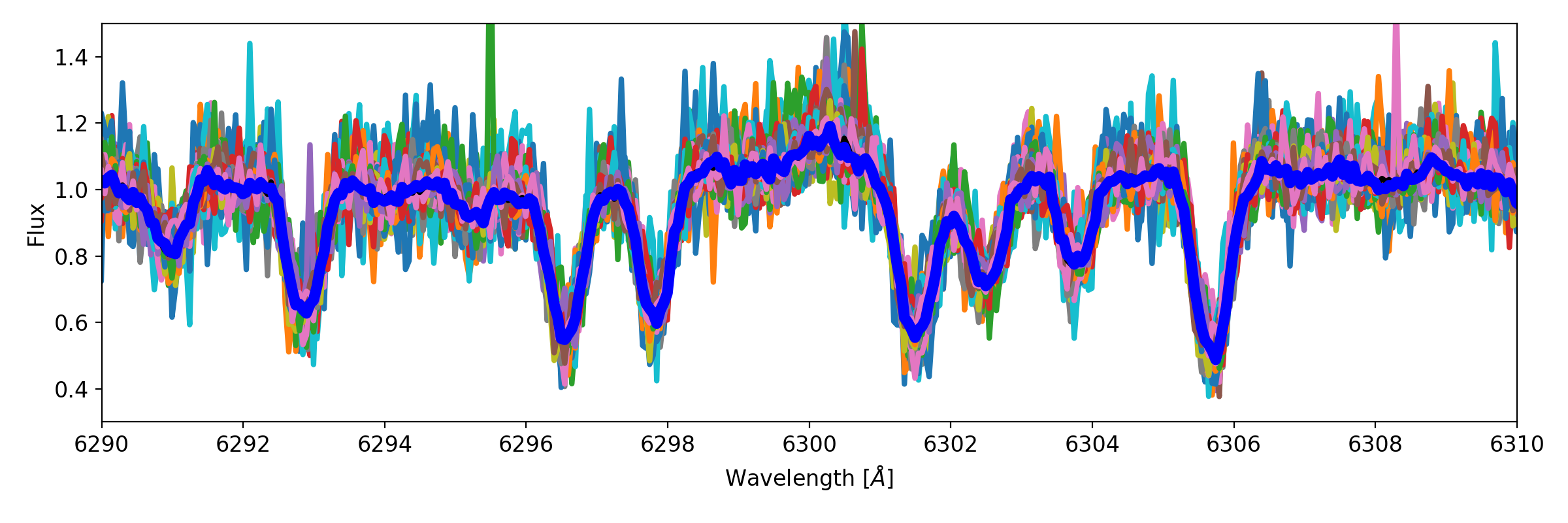}
	\caption{47 HARPS epochs of PDS\,70 shown in background colours, with median combined spectra shown in thick blue. Flux is scaled to the local continuum and centred around the position of  [O\,I] 6300\AA.}
	\label{fig:harps_combine}
\end{figure*}


\section{Observations and Data Reduction}

\label{sec:obs}

PDS\,70 was observed twice with the X-Shooter instrument \citep{2011A&A...536A.105V} on the ESO Very Large Telescope (VLT) in December 2020 and February 2021 (Pr. ID: 105.205R, PI Benisty). These were medium-resolution spectral observations, simultaneously covering three wavelength ranges UV-Blue (UVB) ($\sim$300–560 nm), Visible (VIS) ($\sim$560–1024 nm), and Near-IR (NIR) ($\sim$1020–2480 nm). We note a small variability in the continuum flux levels for each epoch. For further information and data reduction, see Appendix\,\ref{app:xs_obs}.

PDS\,70 was previously observed over multiple seasons by the High Accuracy Radial velocity Planet Searcher \citep[HARPS,][]{2003Msngr.114...20M} on the ESO 3.6\,m telescope at La Silla (Pr. IDs: 098.C-0739, 0104.C-0418, 1101.C-0557, PI Lagrange). HARPS has a high spectral resolution of R = 115,000, with a spectral coverage of 3780 – 6910\AA. Data were reduced by the HARPS pipeline, which removes sky emission using the other fibre.
We then removed telluric absorption features with a developmental version of the molecfit software\footnote{\url{https://support.eso.org/kb/articles/molecfit-experimental-version}}. 
These data included 32 observations from 2018, four from 2019, and 11 from 2020, totalling 47 epochs. A summary of observations is shown in Appendix\,\ref{app:harps_obs}.

To vastly improve the signal-to-noise of the observations, we median combined all 47 HARPS epochs. The continuum normalised region around $\lambda = 6300$ \AA\ is shown in Figure\,\ref{fig:harps_combine}, with this median combined spectra highlighted. Median combining also helped to smooth a few cases of residual noise around the weak sky line that is subtracted from the science spectra.Each epoch is already barycentric corrected by the HARPS pipeline. We measure a low dispersion of radial velocity values across individual observations, with a 1$\sigma$ spread of 0.8\,km/s and typical individual measurement standard errors of 0.03\,km/s. The wavelength values were hence not corrected for individual radial velocities before combination. All subsequent kinematic calculations are adjusted for the average stellar radial velocity value that we determine of 5.5 km/s. Radial velocity measurements and subsequent emission line analysis for this work were carried out using the STAR-MELT Python package \citep{2021MNRAS.507.3331C}\footnote{\url{https://github.com/justyncw/STAR_MELT}}. 


\section{Analysis}
\label{sec:analysis}

\begin{figure*}
\centering
	\includegraphics[width=0.45\textwidth]{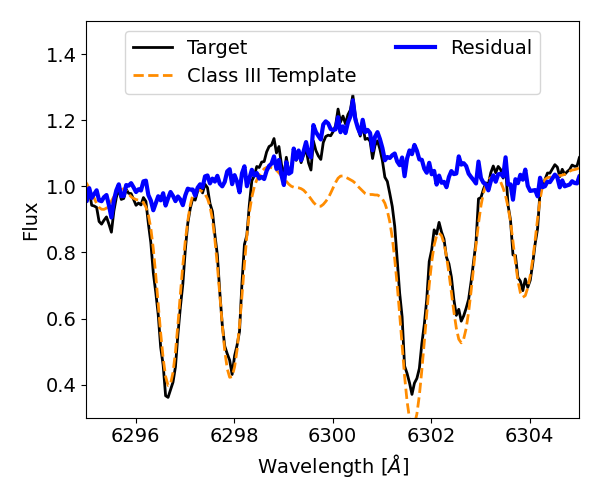}
	\includegraphics[width=0.45\textwidth]{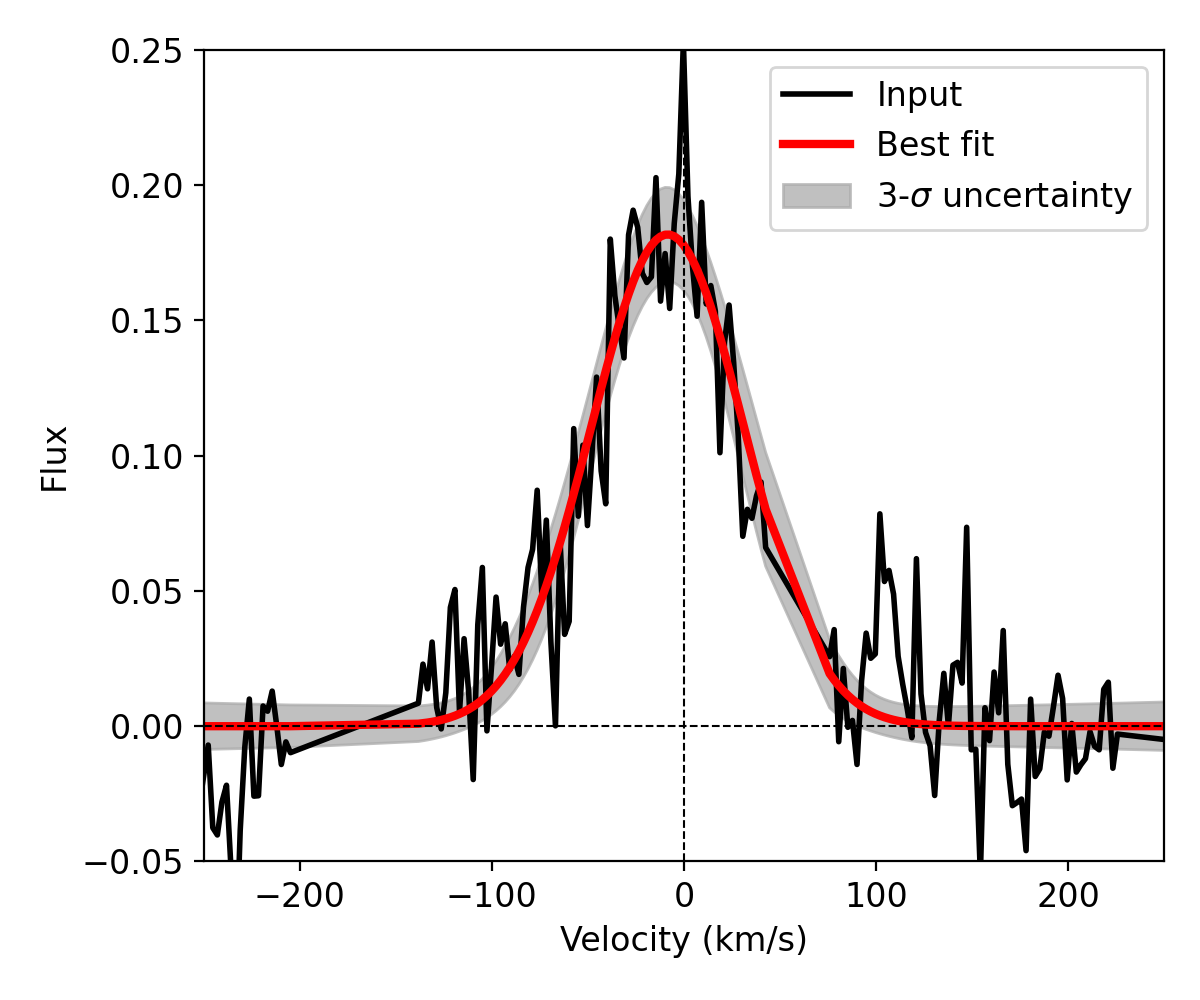}
	\caption{\textit{Left}: Photosphere removal around the [O\,I] 6300\AA\ line of the median combined HARPS spectra (black) with a WTTS template (orange). Resultant residual subtraction of the template from the target is shown in blue. Flux values are scaled to the local continuum. \textit{Right}: Corresponding best-fit model to the continuum flux subtracted [O\,I] line. }
	\label{fig:harps_oi}

\end{figure*}

\subsection{Accretion rate measurements}

We first sought to confirm the accretion rate measurements of PDS\,70 from those previously determined using magnetospheric modelling of the \halpha\ emission \citep{2020ApJ...892...81T}. Using the flux calibrated X-Shooter observations, we attempted a fit following the procedure as described by \citet{2013A&A...558A.114M}, whereby a non-accreting Class III template, reddening and a slab model is used to estimate the observed UV continuum excess, directly resulting from the accretion. However, there is essentially zero UV excess for PDS\,70, so the accretion rates obtained from this method were overestimated. This was apparent from comparison of prominent photospheric absorption lines, such as Ca II $\lambda$ = 423 nm, being too highly veiled in the fit with a slab model. Fitting the PDS\,70 spectrum using only a non-accreting Class III K7 template and no slab did accurately represented such photospheric features. This is due to the fact that accretion in this target, if any, is too low to be detected as continuum excess. Since we detect multiple accretion tracing emission lines in the PDS\,70 spectra, we therefore use the other well studied method of determining the accretion using the line luminosity -- accretion luminosity correlations \citep{2017A&A...600A..20A}.

Emission lines detected in the PDS\,70 X-Shooter and HARPS spectra include those from the hydrogen Balmer series, Ca II H and Ca II K. The Ca II H line is resolved from the adjacent H$\epsilon$ line in the higher-resolution HARPS spectra. We do not detect He I in the optical, nor the Paschen nor Brackett lines in the NIR. The Ca II IR triplet is detected but is deeply embedded in the photospheric absorption lines. Further details of the line flux measurements and determined accretion luminosities are given in Appendix\,\ref{app:Lacc}.

The mean accretion luminosity we derive from the emission lines is log(\lacc/\lsun) $=-2.88 \pm 0.11$. Given the stellar mass \citep[0.76 \msun,][]{2018A&A...617L...2M} and radius \citep[1.26 \rsun,][]{2016MNRAS.461..794P} of PDS\,70, this corresponds to an accretion rate of log(\macc\, yr$^{-1}$) $=-10.06 \pm 0.11$. Since we used the mean HARPS spectra across all epochs, this result is in good agreement with the range of accretion rate values calculated by \citet{2020ApJ...892...81T}. 

\subsection{Wind tracing emission lines}

We performed photospheric subtraction around the potential wind tracing forbidden emission line positions using a Class III template, RXJ1543.1-3920. This template spectra was obtained as part of the PENELLOPE large programme on the ESO Very Large Telescope, using the ESPRESSO instrument \citep[for details of the reduction, see][]{2021A&A...650A.196M}, hence, no spectral degrading was required due to the high-resolution of the template. Following the standard procedure, the target spectra and photospheric template spectra were continuum normalised. The template spectra was then shifted and broadened to respectively match the radial velocity (RV) and projected rotational velocity (\vsini) of the target spectra. For PDS\,70, no absorption line veiling is present. 
The fit resulting in the smallest residuals around the [O\,I] 6300\AA\ line (from both $\chi^2$ calculations and visual inspection) is shown in Figure\,\ref{fig:harps_oi} left. 

Figure\,\ref{fig:harps_oi} right shows the first detection of the resulting [O\,I] 6300\AA\ emission line and best fit model. The signal-noise ratio (SNR) of this detection from the combined HARPS spectra is 9.5. We calculate an EW of $-0.44$\AA, which, together with the X-Shooter continuum flux measurement of $4.7 \pm 0.3 \times 10^{-13}$ erg/s/cm$^2$/nm, gives an integrated line flux of $2.1 \pm 0.4 \times 10^{-14}$ erg/s/cm$^2$. This corresponds to a line luminosity of log($L_{\rm [OI]}$/\lsun)= $-5.08 \pm$ 0.15. We checked whether fewer combined spectra yields the same line luminosity results, and find consistent EWs across each year of observations, but with lower SNRs and higher errors on corresponding model fits. We do also detect the [O\,I] 6300\AA\ line in the X-Shooter spectra, with a SNR of 8 using a median combination of the two epochs. Due to the lower resolution, the photospheric removal results in a poorer subtraction with more prominent residuals either side of the emission, however, we are still able to measure the EW of $-0.46$\AA, providing a line luminosity in good agreement with the HARPS data. The lower resolution of the X-Shooter data is also not as suitable for kinematic line analysis. The remainder of the analysis is hence carried out on the total combined HARPS spectra.

The [O\,I] 6363\AA\ line, which is a factor of 3 weaker than the [O\,I] 6300\AA\ line is within a region of the spectra more significantly affected by photospheric absorption features.We were able to measure the EW of this line above the continuum, finding a value of $-0.16$\AA\ but with a SNR of 3.1. This is consistent with the expected ratio. However, this low SNR does not allow for further analysis of this line. No further forbidden emission lines, such as [S\,II] or [N\,II] were detected in the photospheric removed HARPS spectra, neither was the [O\,I] 5577\AA.  We also note that there is no trace of [Ne\,II], as shown in \citet{2023arXiv230712040P}.

For the best-fit model to the [O\,I] 6300\AA\ emission line, the line intensity is strong enough with respect to the local continuum such that small residual artifacts from the adjacent photospheric removals have negligible effects on the fit. We find consistent fit results when checking fewer combined epochs and subsquent photospheric removals, as with the consistent line luminosity measurements previously noted.
 A single Gaussian low-velocity component (LVC) is adequate to model the line. A linear component was added to the Gaussian component to model the local continuum, as described in \citet{2021MNRAS.507.3331C}. This results in a more accurate fit to the emission component, allowing for slight asymmetries in the overall fit. A combination of broad and narrow component fit could be adopted for this line, but this does not significantly improve the goodness-of-fit of the model, hence is not adopted \citep[following the criteria by][]{2019ApJ...870...76B}. From this best-fit single Gaussian model, the central velocity of the Gaussian component ($V_p$) is -8 $\pm$ 2 km/s (accounting for the standard error of the Gaussian fit, plus the spread in RV values), and the full-width at half-maximum (FWHM) is 89 $\pm$ 5 km/s. The 3$\sigma$ errors of the fit are shown in Figure\,\ref{fig:harps_oi} right.

We also detect the He I 10830\AA\ emission line from each X-Shooter spectrum, which shows blueshifted absorption, indicative of a wind, plus redshifted emission. These observed profiles are strikingly different to the one presented in \citet{2020ApJ...892...81T}, which shows both blue and redshifted absorption components. We checked the alignment between the VIS and NIR arm of the X-Shooter data using overlapping photospheric lines to ensure no velocity offsets, as noted in \citet{2022A&A...666A.188E}. This line and interpretation of the [O\,I] are discussed further in the next section, where we compare properties of the [O\,I] emission to those of other YSOs and the He I profile to the previous observation.


\section{Discussion}
\label{sec:discussion}

With this first detection of [O\,I] emission from PDS\,70, we can compare the measured stellar and line properties to those of other Class\,II YSOs, TDs, and WTTS. Literature data of YSOs from nearby star forming regions were taken from \citet{2014A&A...568A..18M}, for TDs; \citet{2018A&A...609A..87N}, for all Class\,II disc types; \citet{2018ApJ...868...28F}, for TDs; and \citet{2023ApJ...945..112F}, for all disc types. All of the WTTS measurements are taken from this final study and are targets in Upper Sco, hence should also have similar ages to PDS\,70. We include this comparison given its classification as a WTTS based on the \halpha\ profile. However, there are noteworthy discrepancies between both measurements of the emission lines (previously due to spectral resolution, also variability), and the method used to define the WTTS class \citep[e.g., via He I instead of \halpha,][]{2022AJ....163...74T}. Many WTTS still possess low-moderate accretion rates, however, this is difficult to distinguish from chromospheric noise \citep{2013A&A...551A.107M}.

\begin{figure}
	\includegraphics[width=\columnwidth]{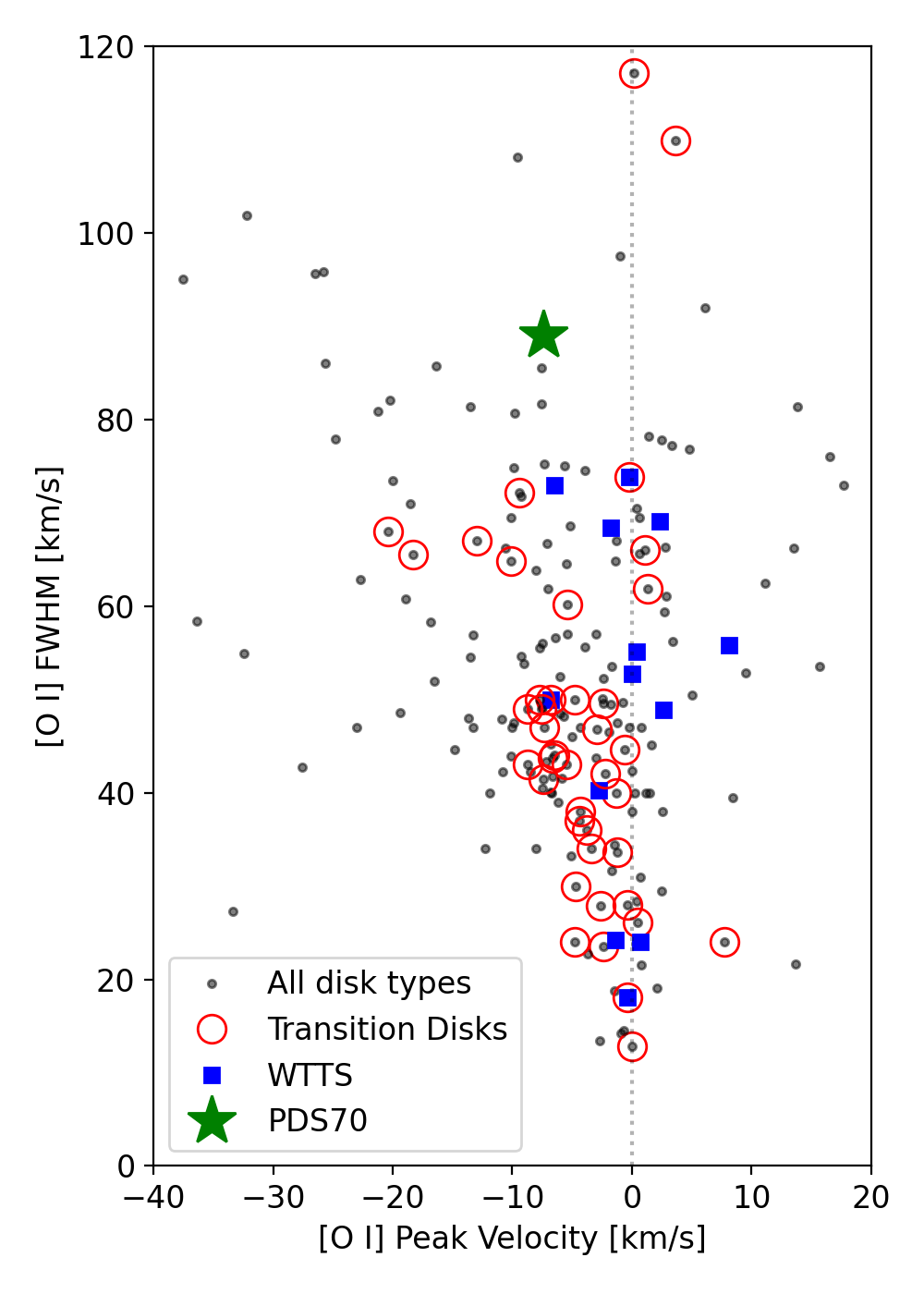}
	\caption{[O\,I] 6300\AA\ peak velocity versus full-width half-maximum comparing the values we obtain for the fit of PDS\,70 to those of other YSOs from the literature values of \citet{2014A&A...568A..18M,2018A&A...609A..87N,2018ApJ...868...28F,2023ApJ...945..112F}. Points show Class II sources, with TDs indicated by the red circles, blue squares show WTTS from Upper Sco with detected [O\,I]. The values we obtained from the best fit of the PDS\,70 [O\,I] is indicated by the green star.}
	\label{fig:oi_kin}
\end{figure}

\begin{figure*}
\centering
	\includegraphics[width=0.8\textwidth]{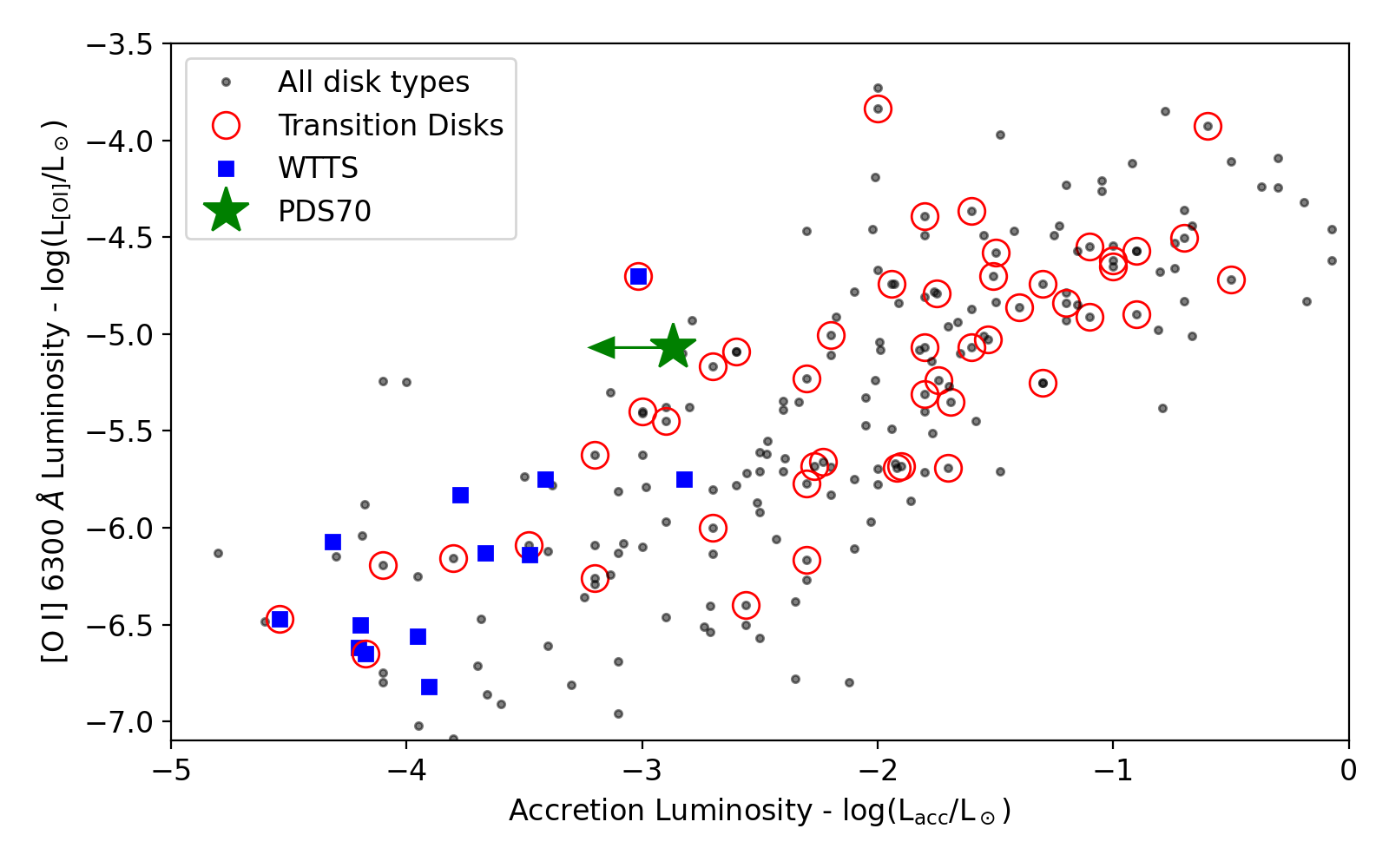}
	\caption{[O\,I] 6300\AA\ line luminosity versus accretion luminosity for PDS\,70 and literature values. The accretion luminosity value of PDS\,70 is as we measure from the average HARPS spectra emission lines. Markers and colours and the same as Figure\,\ref{fig:oi_kin}, as are the sources of literature values.}
	\label{fig:oi_lacc}
\end{figure*}

\subsection{[O\,I] kinematics}
\label{sec:oi_kin}

Figure\,\ref{fig:oi_kin} shows the kinematic values for the fits to the [O\,I] lines; central velocity, $V_p$ and FWHM. For stars with all disc types that have a multicomponent fit to the [O\,I], only the LVC kinematic values are shown. All TDs and WTTS included from previous studies have a single component fit. Typical velocity errors are reported similar to those we measure here, with those from lower-resolutions surveys still $<$ 10 km/s. It is clear from Figure\,\ref{fig:oi_kin} that both the TDs and WTTS occupy a smaller parameter space than other Class II discs. PDS\,70 is clearly an outlier from the WTTS sample.  For the TDs, the mean FWHM is $\sim$ 48 km/s. The value obtained for PDS\,70 is $\sim 2 \sigma$ away. Only two TDs have higher FWHMs \citep[SZ 65 and IM Lup,][]{2018ApJ...868...28F}.

 The fact that the peak is slightly blueshifted is in agreement with models of disc winds \citep{2010MNRAS.406.1553E,2020MNRAS.496..223W,2022EPJP..137.1357E}. Assuming Keplerian rotation ($\Delta v=\sin (i) \sqrt{G M_{\star} / R}$), a disc inclination, $i$, of 50 degrees \citep{2020ApJ...892...81T} and using the FWHM of the profile to approximate the broadening velocity, $\Delta v$ \citep[e.g.,][]{2015ApJ...809..167B,2016ApJ...831..169S,2018ApJ...868...28F}, scaled by the stellar mass \citep[0.76 $\pm$ 0.02 \msun,][]{2018A&A...617L...2M}, we estimate an emitting radius of $\sim$0.1-0.2  AU for the [O\,I]. Since this emitting region is well within the gravitationally bound part of the disc, this also suggests the [O\,I] is tracing a magnetically driven wind rather than photoevaporative. Due to the degeneracies of the Gaussian fitting, it is still possible that two components of the [O\,I] are present, with a narrow component tracing a photoevaporative wind from further out in the disc. However, with this type of composite model, the broad component would be even broader, corresponding to emission from just above the stellar surface and not necessarily the disc. 

Nisini et al (in prep) find a tentative anti-correlation between the Keplerian emitting region of the [O\,I] and the inner cavity size for TDs. This is opposite to what was found in \citet{2019ApJ...870...76B} for the correlation with single component LVC fits and spectral index at 13-31$\micron$, which is used as a proxy for dust in the inner circumstellar disc region. The results from \citet{2019ApJ...870...76B} show that the LVC FWHM decreases as the inferred cavity size increases, however, this assumption was only from the spectral index, and not from direct cavity size measurements. It is possible that once the inner cavity forms, the [O\,I] emission moves inward towards higher density regions of the inner disc as the dust depletion region increases (Nisini et al. in prep). What we find here for PDS\,70 supports this hypothesis, with the emission originating from a higher density inner disc region and not from the inner-edge of the cavity. The absence of further forbidden lines, including no ionised lines, suggests that we are not tracing a photoevaporative wind from the outer cavity wall.

\subsection{[O\,I] intensity}

Figure\,\ref{fig:oi_lacc} shows the comparison between the measured accretion luminosity of the sample of Class II stars versus the line luminosity of the [O\,I] 6300\AA. We include an upper limit to the accretion luminosity measurement of PDS\,70, since we use our measurement from the average combined HARPS spectra, but this may be lower during some phases as previously mentioned. PDS\,70 appears to be an outlier from both the TD and WTTS samples, with high [O\,I] line luminosity for the determined accretion luminosity, suggesting the wind is substantial compared to the infall of accreting material. If the same scaling relations were used, the measured [O\,I] line luminosity of PDS\,70 would correspond to an accretion luminosity of  log(\lacc/\lsun) $\approx$ -1.5, almost two orders of magnitude higher than the accretion luminosity we measure. 

\citet{2023ApJ...945..112F} showed that the discs from Upper Sco have, on average, lower accretion and [O\,I] line luminosities than samples of younger YSOs, but with roughly the same spread in values observed. Hence, PDS\,70 is still an outlier in this regard. Whilst the accretion rate of PDS\,70 is typical for the sample of other TDs, it is clearly at the high end for what is classed as WTTS  \citep{2022AJ....163...74T,2023ApJ...945..112F}. Although, the measured accretion rate when compared to the disc mass of PDS\,70 is low in relation to other YSOs \citep{2019A&A...631L...2M}. Comparing the accretion rate we obtain for PDS\,70 and the cavity size of $\sim$60AU, this agrees with the roughly constant relation from other TDs with sizeable cavities, as shown in \citet{2014A&A...568A..18M}. It is hence the high [O\,I] line luminosity that is setting PDS\,70 apart from the rest of the sample. 

\begin{figure*}
\centering
	\includegraphics[width=0.45\textwidth]{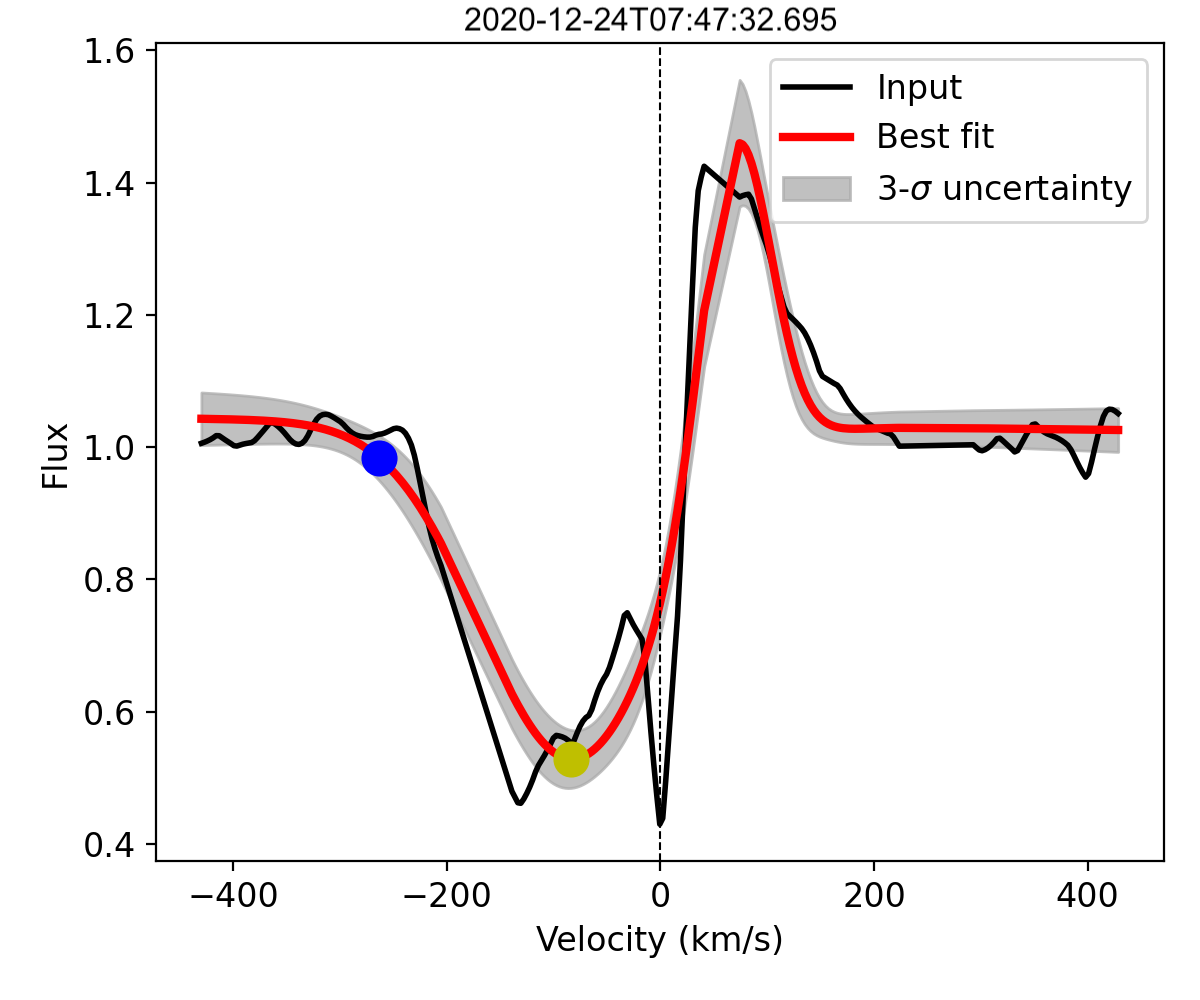}
	\includegraphics[width=0.45\textwidth]{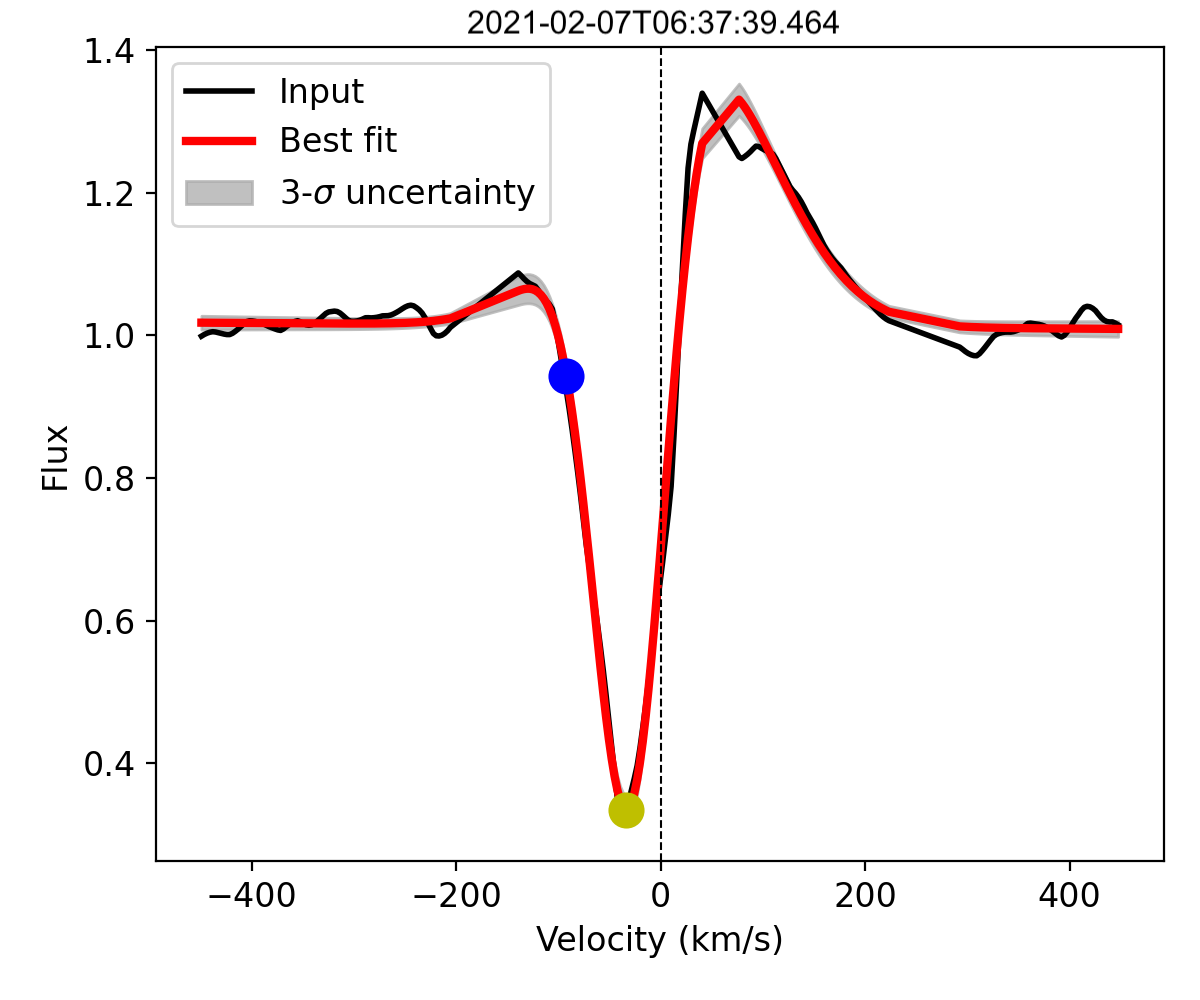}
	\caption{X-Shooter He I 10830\AA\ line profiles for each epoch of PDS\,70 (black). The best composite-model fit is shown in red, with the maximum blueshifted velocity obtained from the 10\% depth of the Gaussian absorption fit shown by the blue circles. The maximum depth of the Gaussian component is shown by the yellow circle. Line profiles are photosphere removed and scaled to the local continuum.}
	\label{fig:he_10830_fits}
\end{figure*}

\citet{2023ApJ...945..112F} note only one `bona-fide' transition disc in their sample from Upper Sco, which happens to be the other WTTS TD outlier in Figure\,\ref{fig:oi_lacc} located above PDS\,70. This target is RXJ1604.3-2130A (hereafter, J1604), which is the focus of many previous studies \citep[e.g.,][]{2018ApJ...868...85P,2020A&A...633A..37S}. J1604 has a misaligned inner disc that casts shadows on the outer disc and is has been a prime candidate for further protoplanet searches, with recent work presenting evidence for a potential companion at the edge of the dust continuum ring \citep{2023A&A...670L...1S}. We see here that it has higher [O\,I] line luminosity than PDS\,70 and a similarly low accretion rate, which has been shown to be highly variable \citep{2020A&A...633A..37S}. It does have a much narrower FWHM for the [O\,I] fit ($\sim20$\,km/s), suggesting a larger Keplerian emitting radius than the inner disc of PDS\,70. The other two TD/WTTS shown in the lower-left of Figure\,\ref{fig:oi_lacc} are 2MASS J16062277-201124 and 2MASS J16151239-2420091. \citet{2022AJ....163...25L}  classified these as TDs from the WISE SEDs but are non- or faint detections from ALMA surveys (J. Carpenter, private communication). These two targets have some of the lowest measurements for both accretion and [O\,I] luminosities and may be at the latest stages of disc evolution.

\subsection{Variable He I 10830\AA\ emission}

The other wind tracing line we detect is He I 10830\AA. Figure\,\ref{fig:he_10830_fits} shows the photosphere subtracted model fits to this line from each of the X-Shooter epochs. The line has a P-Cygni profile, with the redshifted emission located at approximately the same velocity in each epoch. The blueshifted absorption component, however, displays a significant difference in maximum blueshifted velocity and width. This profile is indicative of tracing stellar/disc winds as He I is self absorbed along our line of sight at the outflow velocity corresponding to the maximum blueshifted values. Taking $V_{blue}$ to be 10\% of the maximum depth below the continuum for that Gaussian component \citep[as detailed for inverse P-Cygni profiles and $V_{red}$ in][]{2021MNRAS.507.3331C}, we obtain values of $V_{blue}$ of -277 and -94 km/s for epochs 1 and 2, respectively. These are each below the escape velocity of $\sim$ 480 km/s for PDS\,70.

The profiles of the He I we observe are different from that of the previous detection of this line in \citet{2020ApJ...892...81T}. There, the line has the blueshifted absorption feature, with a measured $V_{blue}$ of $\sim$-85 km/s, and estimated mass loss rate of  $\sim 1 \times 10 ^{-11}$\msun/year, consistent with and MHD inner-disc wind. However, in their previous observation, they detect another absorption feature on the red side of the line, contrary to the redshifted emission we see here. The combination of blue and redshifted absorption is more common for highly accreting CTTS, but uncommon for WTTS stars, with only around 10\% of the WTTS targets in \citet{2023ApJ...944...90T} showing this profile. 

We find that the \halpha\ profiles from the X-Shooter observations do not display the inverse P-Cygni profile that \citet{2020ApJ...892...81T} showed to be variable, in phase with the stellar rotation (see Figure\,\ref{fig:ha_comp}). Hence, the previous observation of the He I line presented there is likely during a phase where this type of profile would be observed in the \halpha, and He is also present in the infalling accretion column. Whilst this kind of double absorption profile may be rare for low- to moderate-accretors, it is likely due to the non-axisymmetric accretion columns along our line of sight to the star and as we see here, a highly variable feature.

\subsection{The peculiarity of PDS\,70}

Detection of this significant inner-disc wind from PDS\,70 would be unusual given its properties even if there were no confirmed protoplanets in the disc. The [O\,I] is broader and brighter than in other WTTS and TDs. The [O\,I] line luminosity is also high for the typical accretion luminosity of WTTS, which is not far in excess of chromospheric lines luminosity. However, with conclusive accretion measures including the further emission line luminosities that we present, concurrent with the magnetospheric modelling accounting for chromospheric emission \citep{2020ApJ...892...81T}, the variable He I profile and the presence of H$_2$ in the inner-disc \citep{2022ApJ...938..134S}, PDS\,70 may in fact be at an earlier stage of disc evolution than previously thought. 

But is it the presence of the planets, having carved out the substantial cavity in the disc, which allows for a high [O\,I] luminosity and inner-disc wind, or is it the disc-wind that facilitate the direct detection of the planets? Although we cannot answer this, recent theoretical modelling work that incorporates MHD winds in conjunction with planets of differing masses show that different combinations result in substructures with varying parameters \citep{2023arXiv230511784W}. Furthermore, the presence of MHD disc winds can influence the formation and migration of planets in the inner disc \citep{2015A&A...579A..65O}. Disc winds have also been observed to be modulated by orbital motions of companions \citep{2014A&A...570A.118F}. Since we show that the inner-disc wind of PDS\,70 is likely MHD in origin (Sec.\,\ref{sec:oi_kin}), further work on the interplay between protoplanets and such winds will be fundamental in untangling the sources of disc substructures. Forthcoming theoretical predictions and synthetic observations may allow for more robust links between the forbidden emission we detect and the physical conditions producing it. Since the search for protoplanets in this early stage of disc evolution is still ongoing, it would be worth focusing efforts on targets that have similar disc-wind properties as PDS\,70.


\section{Conclusions}
\label{sec:conclusions}

We present here the first detection of forbidden emission from the inner-disc of PDS\,70. After photospheric removal, we fit the [O\,I] 6300\AA\ line using STAR-MELT, and characterise its properties to compare to further Class II stars. Kinematic analysis of the line shows that it originates from a radius of $\sim$0.1-0.2\,AU, suggestive of a magnetically driven inner-disc wind, which is supported by the blueshifted peak velocity. The luminosity of the [O\,I] is high for the measured accretion luminosity, and an outlier when compared to other WTTS and TDs. We also show that the He I 10830\AA\ line is highly variable, indicative of both winds and rotating non-axisymmetric accretion flows. We confirm the accretion rate presented in the literature using a different method, and determine log(\macc\, yr$^{-1}$) $=-10.06 \pm 0.11$ from a selection of accretion tracing emission line luminosities. 

 We find that PDS\,70 still has ongoing accretion from the inner-disc, even with no continuum excess observed at UV wavelengths in the X-Shooter observations. The results we find for the substantial inner-disc wind from PDS\,70 suggest that it is MHD in origin, and in combination with the dense inner-disc is shielding the planets and the outer-disc from the photoionisation of the central star that was previously inferred from XUV observations. We do not find direct evidence of a photoevaporative wind from either the inner- or outer-disc. It may be that the significant MHD wind helped to clear out the cavity that was carved by the giant protoplanets, and may have facilitated their direct detections. A similar mechanism could be in play in J1604, allowing the shadows from the inner-disc to be cast on the outer disc. It may also be that the enhanced [O\,I] luminosity and broad profile is the result of the protoplanets significant influence in the disc. Future modelling work on disentangling the effects of planets and winds may help to reconcile these and future observations as we search for further protoplanets around young stars.


\begin{acknowledgements}

We thank the anonymous referee for their report that helped to improve this manuscript.
Funded by the European Union (ERC, WANDA, 101039452). Views and opinions expressed are however those of the author(s) only and do not necessarily reflect those of the European Union or the European Research Council Executive Agency. Neither the European Union nor the granting authority can be held responsible for them. This project has received funding from the European Research Council (ERC) under the European Union's Horizon 2020 research and innovation programme (PROTOPLANETS, grant agreement No.~101002188). RC was partly funded by the Deutsche Forschungsgemeinschaft (DFG, German Research Foundation) in the framework of the YTTHACA Project 469334657 under the project code MA 8447/1-1. AF thanks the support of the Istituto Nazionale di Astrofisica (INAF) through the Large-Grant YODA (YSOs Outflow, Disks and Accretion).

\end{acknowledgements}

\bibliography{bibliography.bib}

\appendix
\section{X-Shooter Observations}
\label{app:xs_obs}

\begin{figure*}
	\includegraphics[width=\textwidth]{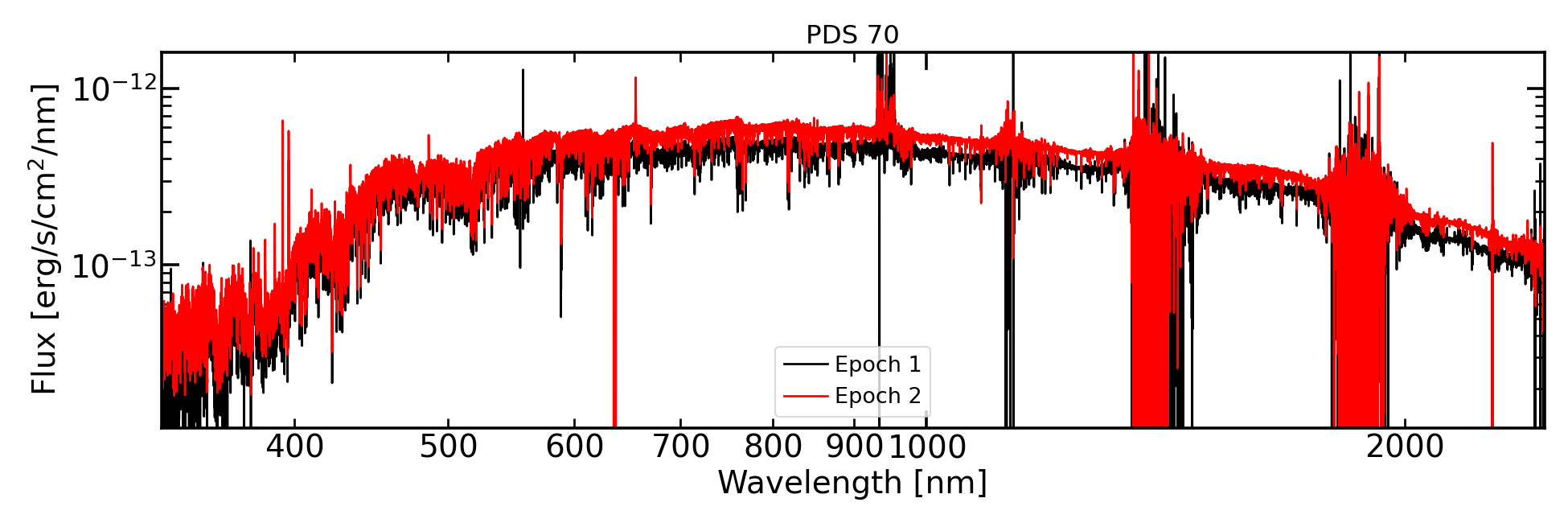}
	\caption{X-Shooter spectra of PDS70. Epoch 1: 2020-12-24T07:47:32.695. Epoch 2: 2021-02-07T06:37:39.464. }
	\label{fig:xs_pds70}
\end{figure*}

\begin{figure}
\centering
	\includegraphics[width=0.7\textwidth]{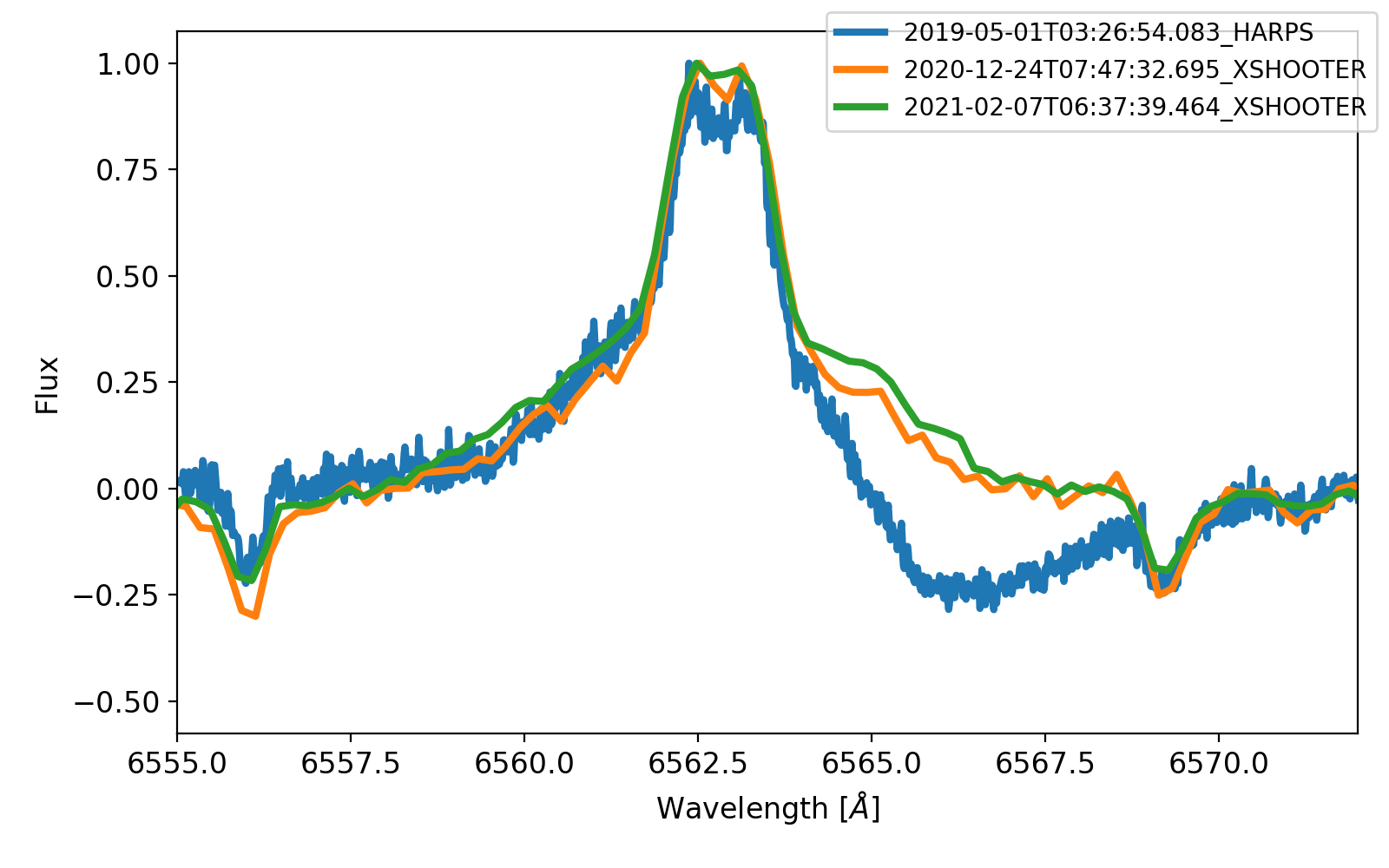}
	\caption{X-Shooter and an example HARPS spectra of PDS70 \halpha\ profile. Flux is normalised and continuum subtracted.}
	\label{fig:ha_comp}
\end{figure}

Figure\,\ref{fig:xs_pds70} shows the two X-Shooter observations of PDS\,70, combining each wavelength range, UV-blue ($\sim$300–560 nm), visible  ($\sim$560–1024 nm), and near-IR ($\sim$1020–2480 nm). Observations were taken in service mode on 24th December 2020 and 7th February 2021, each with clear conditions and seeing $<$2\arcsec . These used a combination of short exposures (45-108\,s) with the wide-slit setups (5.0\arcsec) to measure absolute fluxes, and longer exposures (320-600s) with nod-on-slit integration, with slit-widths of 1.0\arcsec/0.4\arcsec/0.4\arcsec, achieving spectral resolutions of R $\sim$ 5400, 18400, and 11600 in the three respective wavelength ranges. The spectra were reduced using the ESO-Reflex \citep{2013A&A...559A..96F} X-Shooter pipeline v3.5.0 \citep{2010SPIE.7737E..28M} and telluric lines were removed using molecfit \citep{2015A&A...576A..77S,2015A&A...576A..78K}. The final flux calibration was performed by rescaling the flux of the narrow slit to that of the wide slit, as described by \citep{2021A&A...650A.196M}.

Figure\,\ref{fig:ha_comp} shows the \halpha\ profiles of each X-Shooter observation, as well as a comparison HARPS spectra featuring the inverse P-Cygni type profile not observed in the X-Shooter epochs. Here the flux has been normalised and continuum subtracted for comparison between instruments.

\section{HARPS Observing Log}
\label{app:harps_obs}

\begin{table*}[h]
\caption{Observation log of HARPS exposures of PDS\,70}\label{tab:harps_log}
\centering
\begin{tabular}{ccc|ccc}
UTC & MJD & Exp. Time [s] & UTC & MJD & Exp. Time [s] \\
\hline
\hline
2018-03-29T06:33:21.929 & 58206.273170 & 899 & 2018-04-22T05:25:12.656 & 58230.225841 & 1799 \\
2018-03-29T06:48:53.411 & 58206.283952 & 899 & 2018-04-23T04:50:54.677 & 58231.202022 & 1799 \\
2018-03-29T07:04:24.412 & 58206.294727 & 899 & 2018-05-01T04:29:40.280 & 58239.187272 & 1799 \\
2018-03-29T07:19:55.403 & 58206.305502 & 899 & 2018-05-01T05:00:11.202 & 58239.208463 & 1799 \\
2018-03-29T07:36:39.719 & 58206.317126 & 899 & 2018-05-06T03:28:07.285 & 58244.144529 & 1799 \\
2018-03-29T07:52:10.410 & 58206.327898 & 899 & 2018-05-06T03:58:38.487 & 58244.165723 & 1799 \\
2018-03-30T05:41:37.995 & 58207.237245 & 899 & 2018-05-13T05:09:56.831 & 58251.215241 & 1799 \\
2018-03-30T05:57:09.426 & 58207.248026 & 899 & 2018-05-13T05:40:27.781 & 58251.236433 & 1799 \\
2018-03-30T06:12:40.418 & 58207.258801 & 899 & 2019-02-13T08:32:20.542 & 58527.355793 & 1799 \\
2018-03-30T08:22:46.841 & 58207.349153 & 899 & 2019-02-13T09:02:50.943 & 58527.376979 & 1799 \\
2018-03-30T08:38:17.403 & 58207.359924 & 899 & 2019-05-01T03:26:54.083 & 58604.143682 & 2398 \\
2018-03-30T08:53:48.465 & 58207.370700 & 899 & 2019-05-01T04:07:23.607 & 58604.171801 & 2398 \\
2018-03-31T03:39:16.712 & 58208.152277 & 899 & 2020-02-27T05:23:41.660 & 58906.224788 & 1799 \\
2018-03-31T03:54:47.424 & 58208.163049 & 899 & 2020-02-29T05:14:52.221 & 58908.218660 & 1799 \\
2018-03-31T06:35:19.070 & 58208.274526 & 899 & 2020-02-29T05:45:23.028 & 58908.239850 & 1799 \\
2018-03-31T06:50:49.412 & 58208.285294 & 899 & 2020-03-12T05:12:57.551 & 58920.217333 & 1799 \\
2018-03-31T08:28:24.178 & 58208.353058 & 899 & 2020-03-12T06:02:39.146 & 58920.251842 & 1799 \\
2018-03-31T08:43:55.479 & 58208.363837 & 899 & 2020-03-13T04:36:00.448 & 58921.191672 & 1799 \\
2018-04-18T05:12:34.801 & 58226.217069 & 1799 & 2020-03-13T05:06:32.210 & 58921.212873 & 1799 \\
2018-04-19T05:04:01.821 & 58227.211132 & 899 & 2020-03-14T06:32:01.043 & 58922.272234 & 1799 \\
2018-04-19T05:19:33.003 & 58227.221910 & 899 & 2020-03-14T07:02:32.253 & 58922.293429 & 1799 \\
2018-04-20T05:19:03.239 & 58228.221565 & 1799 & 2020-03-15T04:16:16.025 & 58923.177963 & 1799 \\
2018-04-20T05:49:35.891 & 58228.242777 & 1799 & 2020-03-15T04:46:47.164 & 58923.199157 & 1799 \\
2018-04-21T05:51:29.173 & 58229.244088 & 1799 &  &  &  \\

\end{tabular}
\end{table*}

\section{Accretion Luminosity}
\label{app:Lacc}

Accurate measurement of the line fluxes required subtracting the photospheric absorption features from the PDS\,70 spectra. We used a HARPS main sequence (MS) K7 star template, HD35650, since class III templates can still have significant chromospheric emission from these accretion tracing lines. The H$\delta$ line had the most significant photospheric contamination around the emission line, but we were able to adequately remove this to estimate the line flux, albeit with higher uncertainties than the other emission lines. 

We measured the equivalent widths of the continuum normalised, higher resolution HARPS spectra. We took the mean continuum absolute flux values from the two calibrated X-Shooter observations around each line. Multiplying these quantities hence provided absolute integrated flux for each emission line. This allowed for the luminosities of each emission line to be determined, using the distance of 112.4\,pc  \citep[][]{2021A&A...649A...1G}. Line luminosities were then converted to accretion luminosities following the \citet{2017A&A...600A..20A} relations. These results are summarised in Table\,\ref{tab:lacc}.

\begin{table*}
\caption{Flux calculated from median combined photosphere removed HARPS spectra, using the average X-Shooter continuum flux. Corresponding line luminosities are shown with accretion luminosities from \citet{2017A&A...600A..20A} correlations.}\label{tab:lacc}
\centering
\begin{tabular}{l|c|c|c|c|c|c}
    &  H$\alpha$& H$\beta$   & H$\gamma$ & H$\delta$  & Ca II (H)  &  Ca II (K)  \\
    \hline
    \hline
Continuum Flux & 5.25 $\pm$ 0.07  & 3.13 $\pm$ 0.11 & 2.25 $\pm$ 0.12 & 1.21 $\pm$ 0.35 & 0.65 $\pm$ 0.18 & 0.55 $\pm$ 0.11\\

\scriptsize{[10$^{-13}$ erg/s/cm$^{2}$/nm]}  & & & & & &\\

Integrated Flux  & 19.9 $\pm$ 0.01   & 3.91 $\pm$ 0.03 & 1.47 $\pm$ 0.12 & 1.82 $\pm$ 0.55 & 2.07 $\pm$ 0.27 & 3.55 $\pm$ 0.13 \\

\scriptsize{[10$^{-14}$ erg/s/cm$^{2}$]}  & & & & & &\\

Line Luminosity &  -4.10 $\pm$ 0.03   & -4.81 $\pm$ 0.04 & -5.24 $\pm$ 0.08 & -5.14 $\pm$ 0.18 & -5.09 $\pm$ 0.09 & -4.85 $\pm$ 0.06 \\
\scriptsize{log $(L_{line}$/\lsun)}  & & & & & &\\

Accretion Luminosity  & -2.90 $\pm$ 0.06   & -2.90 $\pm$ 0.08 & -3.09 $\pm$ 0.17 & -2.86 $\pm$ 0.28  & -2.74 $\pm$ 0.18 & -2.79 $\pm$ 0.17 \\
\scriptsize{log $(L_{acc}$/\lsun)}  & & & & & &\\

Accretion Rate  & -10.08 $\pm$ 0.11  & -10.08 $\pm$ 0.11 & -10.27 $\pm$ 0.11 & -10.04 $\pm$ 0.11  & -9.92 $\pm$ 0.11 & -9.97 $\pm$ 0.11 \\
\scriptsize{log (\msun/year)}  & & & & & &\\

  \end{tabular}
\end{table*}

\end{document}